\documentclass[
aps,
prd,
10pt,
twocolumn,
nofootinbib,
preprintnumbers,
superscriptaddress,
showpacs,
amsmath,
amssymb,
floatfix]{revtex4-1}

\usepackage{comment}
\usepackage{booktabs,makecell}
\usepackage{graphicx}
\usepackage{amsmath,amssymb}
\usepackage{amsfonts}
\usepackage{xspace} 
\usepackage[usenames]{color}
\usepackage{dcolumn}
\usepackage{bm}
\usepackage{mathrsfs}
\usepackage[colorlinks=true,citecolor=blue]{hyperref}
\usepackage[all]{hypcap} 
\usepackage[utf8]{inputenc} 
\usepackage{slashed}
\usepackage{multirow}
\usepackage{rotating}
\usepackage{nicematrix}
\definecolor{orange}{rgb}{1,0.5,0}
\usepackage{tabularx}

\usepackage[normalem]{ulem}

\def\dd{{\rm d}}
\def\ee{{\,\rm e}}
\def\rrBM{\ifmmode r^{\text{BM}}_{\text{ratio}} \else {$r^{\text{BM}}_{\text{ratio}}$}\fi}
\def\rrDM{\ifmmode r^{\text{DM}}_{\text{ratio}} \else {$r^{\text{DM}}_{\text{ratio}}$}\fi}
\def\X{{\text{X}}}
\def\K{{\text{K}}}
\def\BM{{\text{BM}}}
\def\DM{{\text{DM}}}
\def\Msun{{\text{\,M}_\odot}}
\def\gev{{\text{\,GeV}}}
\def\mev{{\text{\,MeV}}}
\def\hz{{\text{\,Hz}}}
\def\khz{{\text{\,kHz}}}

\def\tot{{\text{tot}}}

\begin{document}

\title{Rapidly spinning dark matter-admixed neutron stars}

\author{Lorenzo Cipriani}\email{lorenzo.cipriani@graduate.univaq.it}
\affiliation{Dipartimento di Scienze Fisiche e Chimiche, Università dell’Aquila, via Vetoio, I-67100, L’Aquila, Italy}
\affiliation{INFN, Laboratori Nazionali del Gran Sasso, I-67100 Assergi (AQ), Italy}

\author{Edoardo Giangrandi}\email{edoardo.giangrandi.1@uni-potsdam.de}
\affiliation{CFisUC, Department of Physics, University of Coimbra, Rua Larga P-3004-516, Coimbra, Portugal}
\affiliation{Institut für Physik und Astronomie, Universität Potsdam, Karl-Liebknecht-Str.24-25, Potsdam, Germany}

\author{Violetta Sagun}\email{v.sagun@soton.ac.uk}
\affiliation{Mathematical Sciences and STAG Research Centre, University of Southampton, Southampton SO17 1BJ, United Kingdom}

\author{Daniela D. Doneva}\email{daniela.doneva@uni-tuebingen.de}
\affiliation{Theoretical Astrophysics, Eberhard Karls University of Tübingen, Tübingen 72076, Germany}

\author{Stoytcho S. Yazadjiev}\email{yazad@phys.uni-sofia.bg}
\affiliation{Department of Theoretical Physics, Sofia University ``St. Kliment Ohridski", Sofia 1164, Bulgaria}
\affiliation{Institute of Mathematics and Informatics, Bulgarian Academy of Sciences, Acad. G. Bonchev St. 8, Sofia 1113, Bulgaria}

\begin{abstract}
Millisecond pulsars, representing the older neutron star population, are believed to have undergone a prolonged period of dark matter accumulation, resulting in a higher dark matter content. Their extreme rotation makes them unique laboratories for studying rapidly rotating neutron stars admixed with dark matter. In this work, we model uniformly rotating neutron stars with a dark matter component that rotates independently from the baryon matter, allowing for the investigation of both co-rotating and counter-rotating scenarios. We examine the impact of dark matter rotation on the macroscopic properties of neutron stars, including the mass-radius relation, the mass-shedding Keplerian limit, and moments of inertia, for various dark matter particle masses and total fractions, considering both core and halo distributions. Our findings provide a more comprehensive understanding of how dark matter influences the equilibrium properties of rotating neutron stars, offering new insights into the astrophysical implications of self-interacting dark matter.
\end{abstract}

\maketitle

\section{Introduction}

Neutron stars (NSs) are among the most extreme astrophysical bodies, serving as invaluable natural laboratories for modern physics. While predicted already in 1934~\cite{baade1934super}, their first observation was made possible by the radio pulses of the star today known as PSR B1919+21~\cite{HEWISH1968}. In 1982 the first millisecond pulsar, PSR B1937+21, was discovered~\cite{Backer1982}, characterized by a period of around $1.55\,\text{ms}$; the fastest-spinning pulsars currently known are PSR J1748-2446ad, discovered in 2004~\cite{Hessels:2006ze}, with a period of only $1.4\,\text{ms}$ (716.36 Hz) and 4U 1820–30 (J1820-30A) with a spin frequency of 716 Hz~\cite{Jaisawal_2024}.

Such rapid rotation can be attained as a result of stellar evolution in closely interacting binary systems, whereas the NS not only accretes mass but also gains angular momentum from its non-degenerate companion star due to the conservation of angular momentum. This mechanism is called ``recycling" spin-up~\cite{1991PhR...203....1B}.

The millisecond pulsars have a weak magnetic field, exhibit extremely slow spin-down rates, and can be several billion years old~\cite{1991PhR...203....1B,Wijnands1998,Guillot:2019vqp}. If these long-lived stars reside in regions such as the cores of galaxies or globular clusters, their strong gravitational field could allow them to accrete a significant fraction of dark matter (DM)~\cite{PhysRevD.82.063531,Bell:2020obw}.

In contrast to symmetric DM which undergoes particle-antiparticle annihilation, asymmetric DM is steadily accumulated during the lifetime of the stellar object, forming gravitationally stable configurations. The measurement of heavy pulsars with masses above $2 M_\odot$ and the detection of old NSs~\cite{PhysRevD.89.015010,PhysRevD.40.3221,PhysRevLett.113.191301,PhysRevD.85.023519,PhysRevD.81.123521,PhysRevD.82.063531,LIU2025101740} put a constraint on the DM particle mass and the allowed relative fraction based on the scenario in which accumulated asymmetric DM exceeds the Chandrasekhar mass, leading to gravitational collapse into a black hole and disrupting the star~\cite{Kouvaris:2013awa,Zurek:2013wia}. However, the presence of a repulsive self-interaction is sufficient to prevent this gravitational collapse, which would otherwise occur when a critical amount of DM is accreted.

The concept of self-interacting DM (SIDM) has gained significant interest as a potential solution to several discrepancies where the Lambda cold dark matter ($\Lambda\text{CDM}$) cosmological model falls short~\cite{Suarez:2013iw}. These mismatches are particularly evident when comparing theoretical predictions to observational data for dwarf galaxies~\cite{Moore_1999}. Thus, the core-cusp problem refers to the inconsistency between the observed flat DM density profiles in dwarf galaxies and the steeply rising distributions predicted by simulations~\cite{deBlok:2010}. Another striking case is the well-known Bullet Cluster (1E 0657-56), often cited as one of the strongest pieces of evidence for the particle nature of DM~\cite{Markevitch_2004,2006cosp...36.2655M}, along with other merging galaxy clusters such as Pandora's Cluster (Abell 2744) and MACS J0025.4-1222~\cite{Eckert:2022qia, Clowe_2006, Randall:2008ppe}. However, the estimated collision velocity of the Bullet Cluster, around $1000~\text{km/s}$, appears highly unlikely within the $\Lambda\text{CDM}$ model~\cite{Lee2010}. SIDM in the form of a strongly-coupled, dilute Bose-Einstein condensate would solve these problems and more~\cite{Harko:2011xw,Suarez:2013iw}. Nevertheless, bosonic DM is not the only candidate, self-interacting or non-interacting fermionic DM is a realistic DM model as well~\citep{Nelson:2018xtr,Das:2020ecp}. In the latter case, the Fermi pressure provides stability against gravitational collapse~\citep{Ivanytskyi:2019wxd,Sagun:2021oml}.

In the scenario in which DM is effectively self-interacting and has negligible coupling with particles of the Standard Model, we can consider gravity as the only relevant force at play. This assumption is fully justified by comparing the DM-BM cross-section constraints coming from the on-ground DM direct detection experiments~\cite{XENON:2023cxc} and astrophysical constraints, e.g. the 
Bullet Cluster~\cite{Randall:2008ppe}, with the strength of the nuclear interaction, which is at least 21 orders of magnitude higher. Thus, two gravitationally interacting fluids can coexist within the same gravitational well, forming DM-admixed NSs (DMANSs). This scenario can be realistically realized in two ways~\cite{PhysRevD.102.063028, RafieiKarkevandi:2021hcc,Diedrichs:2023trk,Hajkarim:2024ecp,Rutherford:2024bli,Scordino:2024ehe}. Heavier DM particles ($m_\DM \gtrsim 1 \gev$) tend to condense into a compact core, resulting in a reduction of both the star’s maximum gravitational mass and tidal deformability $\Lambda$ due to the increased gravitational pull the BM feels~\cite{Hippert:2022snq,Liu:2024rix,Koehn:2024gal}. In contrast, lighter DM particles ($m_\DM \sim 100 \mev$) typically form an extended halo around an NS, leading to an increase in the star's outer radius and tidal deformability. The inclusion of rotation introduces another scenario. Since the two components interact only gravitationally, it is expected that DM and BM have different rotational frequencies, it is possible to build a mixed configuration in which the equatorial radius of the baryonic component is bigger than the radius of the DM one while in the polar direction, along the axis of rotation, the opposite happens.

The accumulation of DM within NSs is influenced by factors such as local DM density, scattering cross-sections (i.e. self-interaction cross-section and with BM), and the star’s evolutionary history~\cite{PhysRevD.89.015010,Bell:2020obw,Bell:2020jou}. While conventional accretion models predict low DM fractions—on the order of one-hundredth of a percent~\cite{Ivanytskyi:2019wxd}—alternative scenarios, such as the accretion of DM clumps or the presence of high-density DM subhalos~\cite{Stref:2016uzb}, could lead to significantly higher values. An overdense DM spike is expected at the Galactic center, concentrated around the supermassive black hole~\cite{Lacroix:2018zmg}. However, rapid DM accumulation in such a spike would be most pronounced for NSs located in the central regions of the galaxy. All these processes become particularly important in the case of millisecond pulsars. Rapidly rotating NSs can also form after the merger of two NSs. The long lifespans of both cases provide sufficient time for interactions with surrounding DM, leading to potentially substantial DM fraction. Over such extended periods, even feeble DM-BM interactions or low local DM densities could result in non-negligible DM accumulation.

Studying rapidly rotating DMANSs is especially interesting in the framework of the X-ray observations by the NICER telescope. Thus, as it was shown in Refs.~\cite{Miao:2022rqj,Rutherford:2022xeb,Shakeri:2022dwg,Shawqi:2024jmk,Liu:2024swd} DM could affect the x-ray pulsar pulse profiles, and consequently the mass-radius constraints inferred from the NICER observations. 

The properties of non-rotating DMANSs have been thoroughly studied for both isolated and binary configurations~\cite{Bezares:2019jcb,Bauswein:2020kor,DiGiovanni:2022mkn,Emma:2022xjs,Ruter:2023uzc} and the first results have been published for the rotating case~\cite{Routaray:2024lni,Cronin:2023xzc,Konstantinou:2024ynd, Mourelle:2024qgo}. In this work, we investigate the impact of rotation on the macroscopic properties of DMANSs. Using an in-house developed extended version of the \texttt{RNS} code \cite{Stergioulas:1994ea, Stergioulas:2003ep}, we investigate scenarios with varying DM fractions, particle masses, and rotation. We analyze both core and halo configurations of DM, focusing on their influence on the maximum mass and rotational properties of NSs.

In contrast to~\cite{Routaray:2024lni,Cronin:2023xzc}, we do not restrict our analysis to the slowly rotating limit, allowing us to explore configurations up to the mass-shedding limit. This broader approach enables a more comprehensive examination of the parameter space, capturing a wider range of astrophysical scenarios. Moreover, unlike~\cite{Konstantinou:2024ynd}, we extend beyond core configurations, encompassing structures from small, compact cores to very large halos. In addition, we systematically explore the maximally rotating, Kepler limit, models.

The paper is organized as follows. Section~\ref{sec:EoSs} presents the equations of state (EoSs) for both BM and DM, while Section~\ref{sec:NumFrame} outlines the framework for constructing rotating DMANS models along with the modifications implemented in the \texttt{RNS} code. In Section~\ref{sec:results}, we analyze representative configurations and discuss their characteristics. Finally, Section~\ref{sec:conclusions} provides a summary of our findings and future perspectives. We use, unless otherwise specified, natural units in which $\hslash = c = G = 1$.

\section{Baryonic and dark matter EoS}
\label{sec:EoSs}

\subsection{Baryonic matter EoS}
\label{sec:BMEOS}

The properties of hadronic matter are modeled employing the relativistic density functional DD2npY-T EoS~\cite{Shahrbaf:2022upc}. This EoS accounts for both nucleonic and hyperonic degrees of freedom and satisfies several key observational constraints. Specifically, it is consistent with the maximum mass requirement for NSs~\cite{Antoniadis:2013pzd,Romani:2021xmb,Romani:2022jhd}, the tidal deformability constraints inferred from the LIGO/Virgo observations of the GW170817 and GW190425 binary NS mergers~\cite{LIGOScientific:2018cki, LIGOScientific:2020aai}, and the NICER measurements of pulsars~\cite{Miller:2019cac,Riley:2019yda,Miller:2021qha,Riley:2021pdl,Choudhury:2024xbk}. For densities below the nuclear saturation, the EoS is complemented by a crust model based on the generalized relativistic density functional~\cite{Typel:2018wmm}. This crust model describes nuclei arranged in a body-centered cubic lattice surrounded by a uniform electron background, with an additional neutron gas present above the neutron drip density. The transition between the crust and the DD2npY-T EoS is implemented consistently within a unified framework.

To address the effect of BM uncertainties at high densities, in Section~\ref{sec:resultsDD2noY}, we perform calculations for two additional EoSs. The first, DD2~\cite{PhysRevC.81.015803}, includes n, p, e, and $\mu$ degrees of freedom. Compared to the DD2npY-T EoS, which includes hyperons, the DD2 EoS is significantly stiffer and allows for a maximum NS mass of $2.4\Msun$. 

Finally, we consider the Induced Surface Tension (IST) EoS which incorporates the hard-core repulsion fitted to the hadron yields measured in a wide range of the center of mass heavy-ion collisions energies~\cite{Sagun:2017eye}, the long-term attraction and asymmetry energy formulated in agreement with the NSs observations and tidal deformability from the GW170817 binary NS merger~\cite{Sagun:2018cpi,Sagun:2020qvc}. The IST EoS accurately models the nuclear liquid-gas phase transition, including its critical endpoint~\cite{Sagun:2013moa}. In the present work, we utilize the set B of the IST EoS that includes n, p, and e degrees of freedom and was developed in~\cite{Sagun:2020qvc}. 

Those three EoSs, i.e. DD2npY-T, DD2, and IST, were chosen to represent different classes of models with varying nuclear matter properties at the saturation density, particle compositions, and stiffness.

\subsection{Dark matter EoS}
\label{sec:DMEOS}

The model of DM employed in this work describes self-interacting bosons. The dynamics is governed by the Lagrangian~\cite{Colpi:1986ye}
\begin{equation}\label{eq:DMLagr}
    \mathcal{L} = \frac{1}{2}\partial_\nu \phi^* \partial^\nu \phi - \frac{m_{\DM}^2}{2} \phi^*\phi - \frac{\lambda}{4} \left(\phi^*\phi\right)^2\,,
\end{equation}
where $\phi$ is a complex scalar field, $m_{\DM}$ is the mass of the DM particle and $\lambda$ is a dimensionless coupling constant.

At sufficiently low temperatures this scalar field forms a Bose-Einstein condensate, allowing stable configurations to exist. A detailed derivation of the EoS in the total condensation limit, corresponding to the zero-temperature regime, is provided in~\cite{Karkevandi:2021ygv}. Introducing the chemical potential $\mu_{\DM}$, the relevant quantities are given by
\begin{subequations}\label{eq:DMEoSAll}
\begin{align}
    n_{\DM} &= \frac{\mu_{\DM}}{\lambda}\left(\mu_{\DM}^2 - m_{\DM}^2\right)\,,\\
    P_{\DM} &= \frac{1}{4 \lambda}\left(\mu_{\DM}^2 - m_{\DM}^2\right)^2\,,\\
    \varepsilon_{\DM} &= 3P_{\DM} + 2 m_{\DM}^2 \sqrt{\frac{P_{\DM}}{\lambda}}\,,\\
    h_{\DM} &= \log{\frac{\mu_{\DM}}{m_{\DM}}}\,,
\end{align}
\end{subequations}
representing, respectively, the DM number density, pressure, energy density, and specific enthalpy.

\section{Numerical framework}
\label{sec:NumFrame}

\subsection{Equilibrium and structure equations}
\label{sec:NumFrameStructEq}

Two ideal fluids in hydrostatic equilibrium can be described within a simple model when they uniformly rotate along the same axis and interactions among them are neglected. Writing the total energy-momentum tensor as
\begin{equation}\label{eq:tmunu}
    T^{\mu\nu}_{\tot} = T^{\mu\nu}_{\BM} + T^{\mu\nu}_{\DM}\,,
\end{equation}
that is to say, ignoring energy transfer between the two fluids, the continuity equation $\nabla_\mu T^{\mu\nu}_{\tot} = 0$ implies that both $\nabla_\mu T^{\mu\nu}_{\BM} = 0$ and $\nabla_\mu T^{\mu\nu}_{\DM} = 0$. For non-rotating configurations it is easy to recover a generalized form of the Tolman–Oppenheimer–Volkoff (TOV) equations~\cite{PhysRev.55.364,PhysRev.55.374}:
\begin{subequations}\label{eq:TOV}
\begin{align}
    \frac{\dd P_{\BM}}{\dd r} &= -\frac{(\varepsilon_\BM + P_\BM)(M_\tot + 4\pi r^3P_\tot)}{r^2 (1 - \frac{2 M_\tot}{r})}\, ,\\
    \frac{\dd P_{\DM}}{\dd r} &= - \frac{(\varepsilon_\DM + P_\DM)(M_\tot + 4\pi r^3P_\tot)}{r^2 (1 - \frac{2 M_\tot}{r})}\, ,\\
    \frac{\dd M_{\BM}}{\dd r} &= 4\pi r^2 \varepsilon_\BM\, ,\label{eq:mbmTOV}\\
    \frac{\dd M_{\DM}}{\dd r} &= 4\pi r^2 \varepsilon_\DM\label{eq:mdmTOV}\, ,
\end{align}
\end{subequations}
where $M_{\tot} = M_{\BM}+M_{\DM}$ and similarly $P_{\tot} = P_{\BM}+P_{\DM}$. 

Masses of DM and BM star components can be also defined as an integral over the entire stellar volume of $\sqrt{-g} (-2 T^0_0 + T^\mu_\mu)$, which can be computed for each fluid as
\begin{equation}
    M_\X = 4\pi \int_0^\infty\!\!\!\!\! \dd r \, r^2 \int_0^{\pi/2}\!\!\!\!\!  \dd \theta \sin\theta \, \ee^{2\alpha + \gamma} \left( \varepsilon_\X + 3 P_\X\right), \label{eq:mx_NonRot}
\end{equation}
where we indicate with the subscript X either the BM or the DM and the metric fields $\alpha$ and $\gamma$ are defined below in Eq.~\eqref{eq:metric}.
In general, it is true that the total mass $M_\tot$, obtained either from Eqs.~\eqref{eq:mbmTOV} and \eqref{eq:mdmTOV} or through Eq.~\eqref{eq:mx_NonRot}, is equivalent since this is an invariant quantity for the system. However, this is not valid for the single components.

These equations can be solved simultaneously by employing the standard Runge-Kutta method. Initial conditions are specified at $r=0$ imposing zero mass and central pressures interpolated from the respective EoS for a given central energy density. 
The metric fields can finally be reconstructed from the matter fields and complete the construction of a spherically symmetric spacetime.

The radii of each of the components are found using the zero-pressure condition at the surface. At the same time, the BM and DM distributions defined by the central energy density or chemical potential values scale proportionally. As was shown in~\cite{Ivanytskyi:2019wxd} the values of the chemical potentials of BM and DM are related as
\begin{equation}
\label{eq:TOV2}
    \frac{\dd \ln \mu_\BM}{\dd r}=\frac{\dd \ln \mu_\DM}{\dd r} = -\frac{M_\tot + 4\pi r^3P_\tot}{r^2 (1 - \frac{2 M_\tot}{r})}.
\end{equation}

To consider rotating configurations, the Einstein field equations can be solved in axial symmetry by employing the Komatsu-Eriguchi-Hachisu (KEH) scheme~\cite{Komatsu:1989zz}, incorporating the modifications proposed in~\cite{Cook:1992}. In particular we have modified the \texttt{RNS}\footnote{\href{http://www.github.com/cgca/rns}{github.com/cgca/rns}} code~\cite{Stergioulas:1994ea, Stergioulas:2003ep}, straightforwardly generalizing the original algorithm as follows. 

The general line element is expressed in quasi-isotropic coordinates as
\begin{equation}\label{eq:metric}
\begin{split}
    \dd s^2 = &- \ee^{\gamma + \rho}\dd t^2 + \ee^{2\alpha} (\dd r^2 + r^2 \dd \theta^2)\\
     &+ \ee^{\gamma - \rho} r^2 \sin^2\theta(\dd \phi - \omega \dd t)^2\,,
\end{split}
\end{equation}
where $\gamma$, $\rho$, $\alpha$, and $\omega$ represent the metric potentials, each depending on $r$ and $\theta$. The circumferential radius is recovered from $r$ as
\begin{equation}\label{eq:circRad}
    R = r \ee^{\frac{\gamma - \rho}{2}}\,.
\end{equation}

Each fluid's four-velocity $u^\mu_\X$ is defined as
\begin{equation}\label{eq:ualpha}
u^\mu_\X = \frac{\ee^{-(\gamma+\rho)/2}}{\sqrt{1 - v_{\X}^2}} \left(1, 0,0,\Omega_\X\right)\,,
\end{equation}
where $v_\X = (\Omega_\X - \omega) r \sin\theta \ee^{-\rho}$ and $\Omega_\X$ is the fluid angular velocity. The sign of $\Omega_\X$ determines if the fluid rotates counter-clockwise or clockwise. For most of our studies we will assume that both fluids rotate in a counter-clockwise rotation ($\Omega_\X > 0$); in Section~\ref{sec:resultsCounter} we will explore the other interesting case in which the DM fluid counter-rotates with respect to the BM ($\Omega_\DM < 0$).

Eq.~\eqref{eq:ualpha} implies that even if one of the two components has zero angular velocity $\Omega_\X$, the linear velocity $v_\X$ remains nonzero as a result of the frame-dragging effect induced by the other component. 

The 4-velocity $u^\mu_\X$ enters in the energy momentum tensor as
\begin{equation}
    T_\X^{\mu\nu} = (\varepsilon_\X + P_\X) u_\X^\mu u_\X^\nu + P_\X g^{\mu\nu}\, ,
\end{equation}
where $g^{\mu\nu}$ is the inverse metric tensor.

\begin{figure*}
    \centering
    \includegraphics[width=\textwidth]{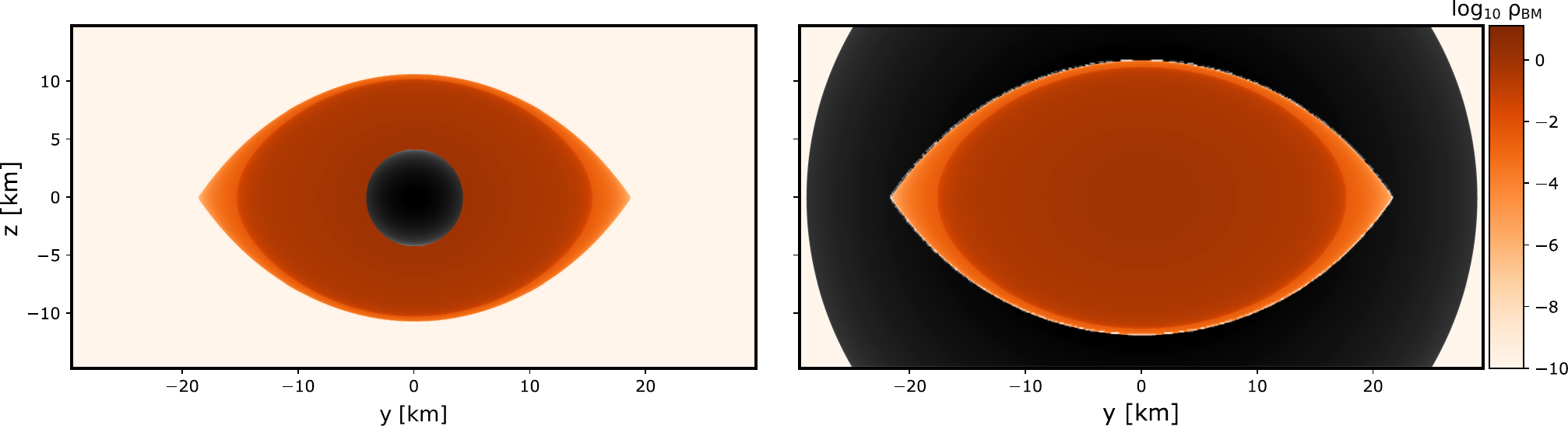}
    \vspace*{-1.5ex}
    \caption{Density profiles in the $y-z$ plane at $x = 0$ for a NS rotating at $\Omega^\BM_\K$, $\Omega_\DM = 0$ and $f_\DM = 5\%$. \textit{Left}: core configuration, $M_\tot = 2\Msun$, $m_\DM = 1\gev$, $\Omega^\BM_\K = 8068 \hz$. \textit{Right}: halo configuration, $M_\tot = 2.08\Msun$, $m_\DM = 250\mev$, $\Omega^\BM_\K = 6841 \hz$.}
    \label{fig:fig1}
\end{figure*}

Studying the non-vanishing Einstein field equations, it is possible to show that the structure of the equations to be solved is analogous to the one-fluid case reported in~\cite{Komatsu:1989zz}. The explicit form of these equations is reported in Appendix~\ref{app:implem}.

The computation of the metric fields, alongside the matter distribution, begins with an initial guess, typically a previous iteration or a non-rotating star. From the initial metric fields, a value of the angular velocity is determined through the first integral of the hydrostationary equilibrium
\begin{equation}\label{eq:hydroEq}
h_\X - \ln u^t_\X = \text{const}\,,
\end{equation}
where $h_\X$ is the specific enthalpy. By coupling Eq.~\eqref{eq:hydroEq} with a prescribed ratio $r^\X_\text{ratio}$ between the polar radius ($r^\X_p$) and equatorial radius ($r^\X_e$) for each fluid, it is possible to compute

\begin{equation}
\label{eq:Omegax}
\Omega_\X = \omega_e^\X \pm \sqrt{1 - \ee^{\frac{\gamma^\X_p + \rho^\X_p - \gamma^\X_e - \rho^\X_e}{2}}} \frac{\ee^{\rho^\X_e}}{r^\X_e}
\end{equation}
and thus $v_\X$. The metric fields are evaluated at either the equatorial radius of the respective fluid (subscript $e$) or the polar radius (subscript $p$)--e.g.
\begin{align}
    \gamma^\BM_e &= \gamma\left( r = r^\BM_e,\, \theta = \pi/2 \right)\,,\\
    \gamma^\BM_p &= \gamma\left( r = r^\BM_p,\, \theta = 0\right)\,.
\end{align}
Note that when $\Omega_\X < 0$ the square root in Eq.~\eqref{eq:Omegax} acquires the negative sign.

The specific enthalpy distribution is computed using
\begin{equation}\label{eq:enthDist}
h_\X = \frac{1}{2} \left[ \gamma^\X_p + \rho^\X_p - \gamma - \rho - \ln(1 - v_\X^2) \right] \,.
\end{equation}
The matter distributions, i.e. the energy densities $\varepsilon_\X$ and pressures $P_\X$, are computed from Eq.~\eqref{eq:enthDist} starting from a fixed value of the central energy density $\varepsilon^\X_c$. The metric potentials are then recalculated, and this iterative process is repeated until convergence is achieved.

Of particular importance for our studies is the choice of a definition for the DM fraction
\begin{equation}\label{eq:fdm}
    f_\DM = \frac{M_\DM}{M_\tot}\,.
\end{equation}
It is well known~\cite{Misner:1973prb} that in general relativity there is no unique definition of an object's mass. In the previous paragraph, we discussed the different mass definitions of the non-rotating case. While Eq.~\eqref{eq:mx_NonRot} can be readily generalized to rotating stars, there is no straightforward rotating version of Eqs.~\eqref{eq:mbmTOV} and \eqref{eq:mdmTOV}. Thus, for rotating stars, we define the DM and BM masses as     
\begin{equation}\label{eq:MgravRot}
\begin{split}
M_\X &= 4\pi \int_0^\infty\!\!\!\!\! \dd r \, r^2 \int_0^{\pi/2}\!\!\!\!\!  \dd \theta \sin\theta \, \ee^{2\alpha + \gamma} \times\\
&\left[ \frac{\varepsilon_\X + P_\X}{1 - v_\X^2} \left(1 + v_\X^2 + 2 \omega r v_\X \ee^{-\rho} \sin\theta \right)+ 2 P_\X \right]\, ,
\end{split}
\end{equation}
and of course, the total mass is the sum of the two.

The rest mass of each component $M^0_\X$ can also be calculated using an integral throughout the star
\begin{equation}
M^0_\X = 4\pi \int_0^\infty \!\!\!\dd r \int_0^{\pi/2} \!\!\!\dd \theta \sqrt{-g} W \rho^0_\X\,,
\end{equation}
where $W = 1 / \sqrt{1 - v_\X^2}$ denotes the Lorentz factor of the fluid element, and $\rho^0_\X = \left(\varepsilon_\X + P_\X\right) \ee^{-h_\X}$ represents the rest mass density. Naturally, the total rest mass of the star is the sum of the two components.

Another important parameter of the rotating stars is the Keplerian angular velocity $\Omega_\K$. It represents the angular velocity of a particle on a circular orbit around the star's equator at zero altitude and is the maximum angular velocity allowed for stable configurations. For the metric element given in Eq.~\eqref{eq:metric} it can be expressed as  
\begin{equation}\label{eq:Ok}
    \Omega^\X_\K = \omega^\X_e + \frac{v_e^\X}{r^\X_e} \ee^{\rho^\X_e}\,,
\end{equation}
where $v_e^\X$ is the equatorial orbital velocity measured by an observer with zero angular momentum in the $\phi$-direction. $\Omega^\X_\K$, in general, depends parametrically on both the central energy density of the fluid X and its angular velocity $\Omega_\X$. Note that Eq.~\eqref{eq:Ok} can be used to compute the Keplerian angular velocity regardless of the configuration we choose for the DM since the velocity of a particle in circular orbit on the equator depends only on the metric fields:
\begin{equation}
\begin{split}
    v_e(r&) =
\frac{e^{-\rho} r^2 \partial_r \omega}{2 + r \left( \partial_r \gamma - \partial_r\rho \right)} \pm \\
& \sqrt{\frac{r \big(\partial_r \gamma + \partial_r \rho\big)}{2 + r \left( \partial_r \gamma - \partial_r \rho \right)} +\frac{ e^{-2\rho} r^4 (\partial_r \omega)^2}{\left[2 + r \left(\partial_r \gamma - \partial_r \rho \right)\right]^2} }\,.
\end{split}
\end{equation}
As before, the minus sign is acquired when the fluid X is counter-rotating.

Another important quantity to study for rotating NSs is their moment of inertia, defined as
\begin{equation}\label{eq:moi}
    I_\X = \frac{J_\BM + J_\DM}{\Omega_\X}\,,
\end{equation}
where the angular momentum $J_\X$ is computed from
\begin{equation}
\begin{split}
    J_\X &= \int \dd^3 x\, \sqrt{-g}\, {{T_\X}^0}_\phi\\
    &= 4\pi\int_0^\infty \!\!\!\!\! \dd r\,r^3\!\! \int_0^{\pi/2}\!\!\!\!\!  \dd \theta \sin^2 \theta \ee^{2\alpha + \gamma -\rho} \left( \varepsilon_\X + P_\X \right) \frac{v_\X}{1 - v_\X^2}\,.
\end{split}
\end{equation}

In the two-fluid system the moment of inertia could be defined for each of the components separately, or the total one~\cite{Leung:2012vea,Cronin:2023xzc}. Since this quantity measures the object's resistance to changes in rotational frequency and depends on the mass distribution of both components, it serves as a potential probe of DM, which will be discussed in the following section.

Interestingly, the core configurations of DM are impressively similar to superfluid NSs where the two fluids correspond to the normal and superfluid  components~\cite{Andersson:2000hy,Prix:2004wq}. This similarity could bring new insights to the DM topic.

\section{Results}\label{sec:results}
In this section, we present the results of our analysis of rotating DMANSs at the mass-shedding limit. Hereafter, as the most representative case, we employ two distinct DM configurations: a DM core obtained for the particle mass $m_\DM = 1\gev$ and a halo for $m_\DM = 250\mev$. Note, that the self-interaction strength value in both cases is fixed to $\lambda = 24\pi$. This choice is arbitrary and made in order to obtain in the non-rotating core limit a maximum mass close to $2\Msun$.

\subsection{Non-rotating dark matter}

Fig.~\ref{fig:fig1} shows a slice of the density profiles for both BM and DM on the $x=0$ plane for two representative configurations: a core (left panel) and a halo (right panel). These models correspond to systems with a total gravitational mass of $2\Msun$ (left) and $2.08\Msun$ (right), a DM fraction of $f_\DM = 5\%$, $\Omega_\DM = 0\hz$, and $\Omega_\BM = \Omega^\BM_K$. Both panels highlight the characteristic cusp that develops at the BM mass-shedding limit because of the very rapid increase of the equatorial radius $R_\BM$ when increasing $\Omega_\BM$~\cite{Paschalidis2017}. We can notice how the deformation due to DM, at least for this particular choice of parameters, is very weak. This is a reflection of the fact that the DM and BM components interact only gravitationally.

\begin{figure}
\includegraphics[width=0.95\columnwidth]{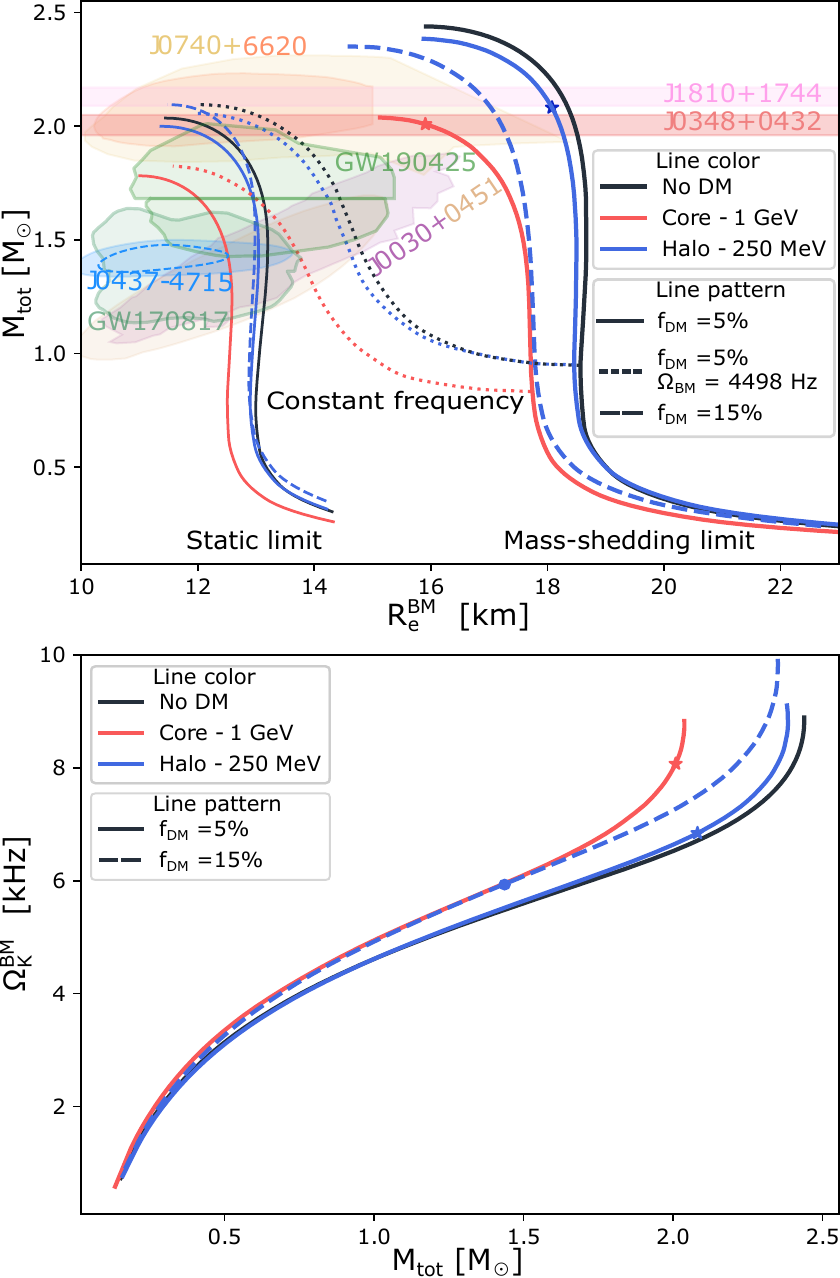}
\vspace*{-1.5ex}
\caption{Sequences of configurations with non-rotating DM $\Omega_\DM = 0$ and two fixed DM fractions $f_\DM=5\%$ (solid lines) and $f_\DM=15\%$ (dashed lines). With red (blue) we indicate the curve that characterizes a core (halo) case with particles of $1\gev$ ($250\mev$). The two star symbols represent the models reported in Fig.~\ref{fig:fig1}. \textit{Top:} The total NS mass as a function of its equatorial radius. Thin lines represent non-rotating BM as a reference. Dotted lines show the sequences rotating at constant angular frequency $\nu_\BM = 716\hz$ (or equivalently an angular velocity $\Omega_\BM = 4498\hz$). 
The pink and light red bands represent the 1$\sigma$ constraints on the mass of PSR J1810+1744~\citep{Romani:2021xmb}, and PSR J0348+0432~\citep{Antoniadis:2013pzd}. The NICER measurement of PSR J0030+0451~\citep{Miller:2019cac,Riley:2019yda} is shown with the purple and yellow contours, the gold and orange contours represent the PSR J0740+6620 measurement~\citep{Miller:2021qha,Riley:2021pdl}, while light blue (95\% CL) and light blue dashed (68\% CL) contours represent the PSR J0437-4715 measurement ~\citep{Choudhury:2024xbk}. LIGO-Virgo detections of 
GW170817~\cite{LIGOScientific:2018cki} and GW190425 \citep{LIGOScientific:2020aai} binary NS mergers are shown in light green. \textit{Bottom:} Mass dependence of the Keplerian angular velocity for DMANS with non-rotating DM (thick lines on the top panel).}
\label{fig:fig2}
\end{figure}

A first analysis of the impact of non-rotating DM on spinning NSs can be carried looking at the top panel of Fig.~\ref{fig:fig2}. Here we report the relation between the total mass given by Eq.~\eqref{eq:MgravRot} and the equatorial radius $R^\BM_e$ for static DMANSs (thin lines) and maximally rotating stars (thick lines). Thin dotted lines between the two depict sequences with fixed rotational rate, namely $f=716$Hz, corresponding to the fastest spinning pulsar as discussed above. Naturally, for low masses, it ends at the Kepler limit sequences. The colors distinguish the different DM configurations: red shows a core configuration, blue is a halo, and finally black is the standard case of no DM. All allowed uniformly rotating stars with $f_\DM = 5\%$ (solid lines) lie in the region bounded by these curves. Finally, the dashed blue lines show another realization of the halo configuration for $f_\DM = 15\%$. Note that such a high DM fraction was chosen to illustrate the effect at its extreme.

Rotation generally tends to increase the maximum mass by approximately $20\%$ and the radius by about $40\%$ for pure BM NSs~\cite{Paschalidis2017}. In the case of a DM core, in the static limit the additional mass present with respect to the $f_\DM = 0$ case mimics a softening of the nucleonic EoS. The effect is a reduction of the maximum mass and radius. Rotation of the BM at the Kepler frequency has the same effect as before, increasing the maximum mass and radius by approximately $15\%$ and $40\%$ respectively. The case of a DM halo exhibits a more peculiar behavior. In the static limit for $f_\DM = 5\%$, we do not observe the expected increase in mass and radius, as predicted by previous studies~\cite{Leung:2011zz,Ellis:2018bkr,Nelson:2018xtr}. This deviation arises from the specific choice of parameters $f_\DM$, $m_\DM$, and $\lambda$. Thus, lighter DM particles of $m_\DM \sim 100\mev$ would form a more extended and dilute halo~\cite{Karkevandi:2021ygv}. Notably, if we keep $m_\DM$ and $\lambda$ fixed while increasing the DM fraction beyond 10\%, the expected growth in mass and radius emerges, as illustrated by the dashed blue line for $f_\DM = 15\%$. When BM rotates at its Kepler angular velocity, the maximum mass and radius for $f_\DM = 5\%$ increase by approximately 20\% and 40\%, respectively.

The DM component, currently assumed to be non-rotating, is only marginally affected by the rotation of the NS, except in the most massive systems. In both core and halo configurations, low-mass systems are associated with a small Keplerian angular velocity $\Omega^\BM_\K$ and a weak frame-dragging potential $\omega$. This results in a minimal deformation of the DM distribution: the equatorial radius $R^\DM_e$ exceeds the polar radius $R^\DM_p$ by less than $0.1\%$. Near the last stable configuration this deformation increases but remains modest, with a maximum difference of approximately 2\%. In contrast, the BM exhibits significantly larger deformations, with the equatorial radius exceeding the polar radius by about 50\% even for low-mass systems. This percentage gradually increases as the system approaches the last stable configuration.

Higher DM fractions partially change this consideration, e.g. the blue dashed line in Fig. \ref{fig:fig2} for a halo configuration with $f_\DM=15\%$. While the deformation from spherical symmetry remains small, the change in $R^\DM_e$ that is found with the current setup is much bigger, leading to the formation of a mixed sequence where at low masses the DM forms a halo. In contrast, at high masses, it is characterized by $R^\BM_e > R^\DM_e$ but $R^\BM_p < R^\DM_p$. This is what happens for the dashed line on the upper panel of Fig.~\ref{fig:fig2}; the switch happens at a mass of $1.44\Msun$. The blue dot highlights the specific configuration marking this transition.

The bottom panel of Fig.~\ref{fig:fig2} shows a comparison between the Kepler frequencies across the three configurations for varying total gravitational masses. Core configurations consistently exhibit higher rotational frequencies compared to both the halo and $f_\DM = 0$ cases. The stronger gravitational pull at the equator requires an increased centrifugal force to reach the mass-shedding limit, and hence a higher angular frequency. Halos show instead much smaller deviations from the pure BM case: only for masses above $1.5\Msun$ there is an appreciable difference between the two. Increasing the DM fraction in this last case has the expected behavior of increasing the maximum angular velocity while maintaining the same qualitative trajectory. A higher DM fraction at the fixed total mass implies a higher DM mass, thus a lower radius and a higher angular velocity.

\begin{figure}
\includegraphics[width=0.95\columnwidth]{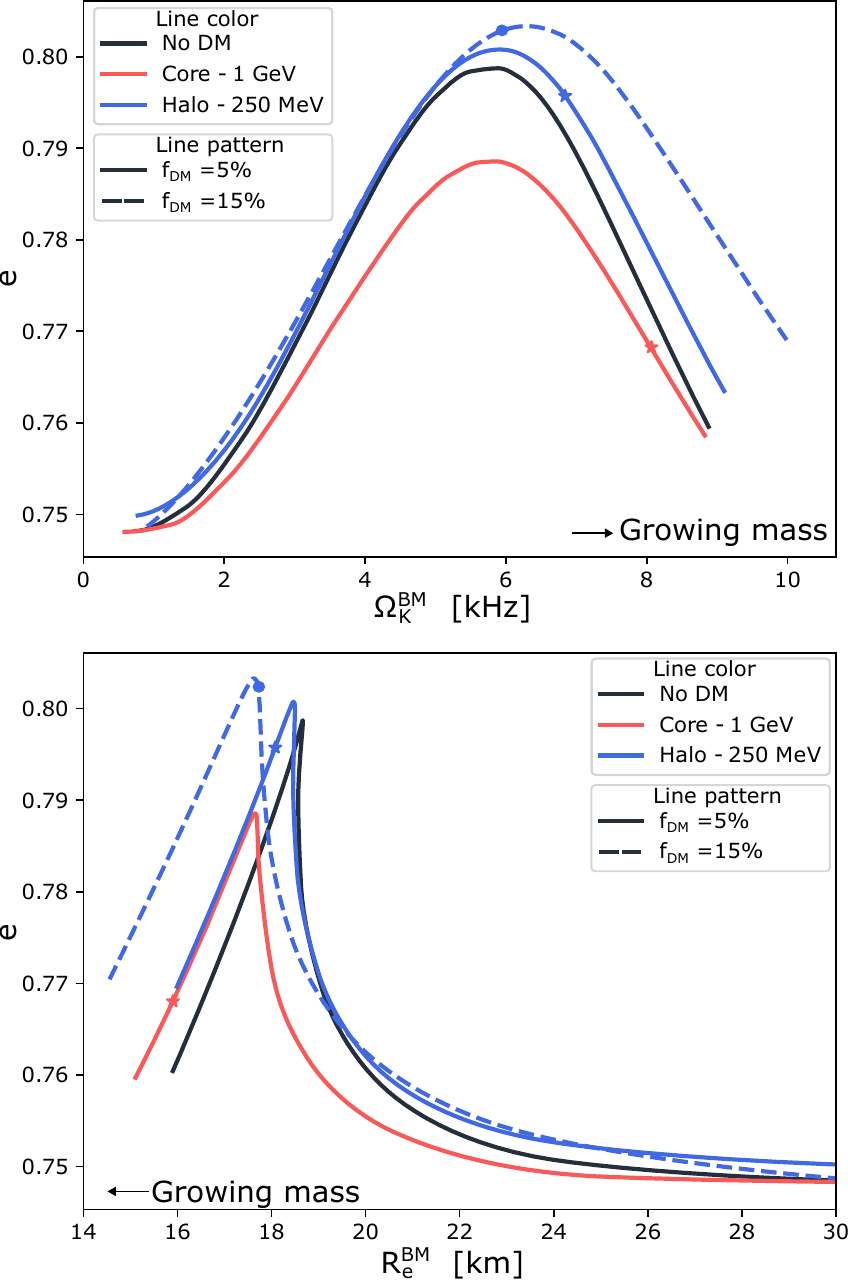}
\vspace*{-1.5ex}
\caption{Both panels show the stable configurations of maximally rotating (Kepler) sequences with varying central energy density $e^\BM_c$. \textit{Top:} Oblateness as a function of the angular velocity $\Omega_\BM = \Omega^\BM_\K$. \textit{Bottom:} Oblateness as a function of the BM equatorial radius $R^\BM_e$.}
\label{fig:fig3}
\end{figure}

Fig.~\ref{fig:fig3} show the two key aspects of the maximally rotating sequences, presenting the oblateness, defined as \begin{equation}
    e = \sqrt{1 - \left(\frac{R^\BM_p}{R^\BM_e}\right)^2}\, .
\end{equation}
The top panel of Fig.~\ref{fig:fig3} shows $e$ as a function of the angular velocity $\Omega_\BM = \Omega^\BM_\K$. All four profiles show similar behavior. Low-mass stars rotating at angular velocities on the order of hundreds of Hz exhibit only a slight increase in oblateness with growing angular velocity $\Omega_\BM$--and hence with the total mass. As $\Omega_\BM$ increases, the oblateness reaches a maximum value. The angular velocity at which this peak occurs is consistent between all configurations: they closely follow the $f_\DM=0$ reference at $\Omega_\BM \approx 6\khz$. Then the oblateness tends to decrease until the last stable configuration is reached. 

The bottom panel of Fig.~\ref{fig:fig3} illustrates a comparison of oblateness against the baryonic equatorial radius of the maximally rotating configurations. At high radii--i.e., low masses and low angular velocities--the configurations show a similar trend towards slowly changing values of the oblateness. It is intriguing to observe that as the systems approach lower radii, halos, and cores become degenerate while decreasing linearly with the radius. They exhibit the same deformation, despite having drastically different masses, angular velocities, and DM configurations.  Specifically, as the difference between the configurations becomes very small, the DM halo has a radius several kilometers larger than the NS, while the DM core is always such that $R^\BM_e \approx 3 R^\DM_e$. Changing the DM fraction shifts the location of this linear regime to lower radii while keeping the slope approximately the same. Note that when the oblateness reaches its maximum the system is already in the mixed configuration characterized by $R^\BM_e > R^\DM_e$ but $R^\DM_p > R^\BM_p$.

\begin{figure}
\includegraphics[width=0.95\columnwidth]{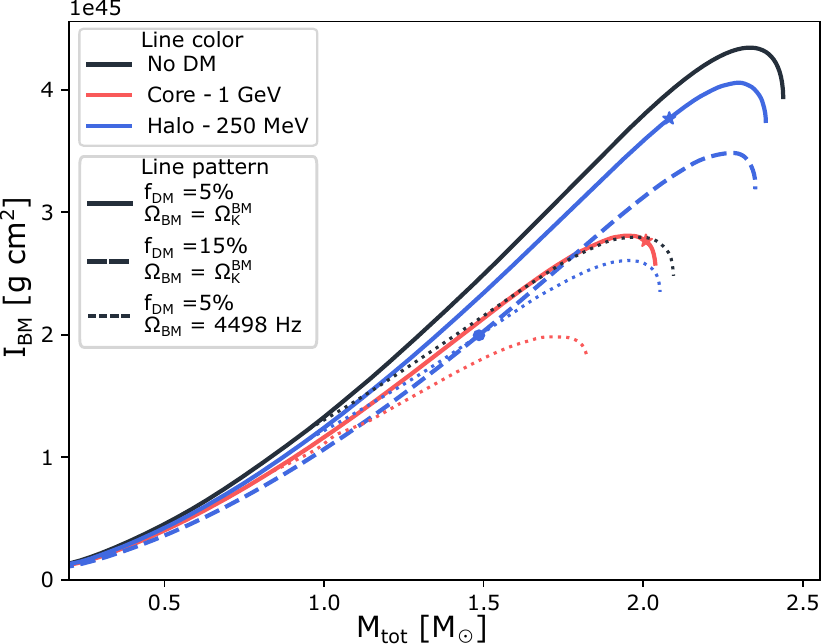}
\vspace*{-1.5ex}
\caption{Moment of inertia for various configurations of DMANSs as a function of the total mass. Here $J_\DM = 0$. Dotted lines illustrate the sequences rotating at constant angular velocity $\Omega_\BM = 4498\hz$ (or equivalently an angular frequency of $\nu_\BM = 716\hz$), while the rest of the lines correspond to maximally rotating configurations.}
\label{fig:fig4}
\end{figure}

Finally, Fig.~\ref{fig:fig4} presents the moment of inertia of BM, $I_\BM$, as defined in Eq.~\eqref{eq:moi}, for the different NSs configurations as a function of the total gravitational mass. Solid lines represent maximally rotating stars with the DM fraction of 5\%, dotted lines represent the sequences of stars at a constant angular frequency of $\nu_\BM = \Omega_\BM / 2\pi = 716\hz$, and the dashed line corresponds to the maximally rotating halo sequence for the 15\% DM fraction. For low-mass stars, the moment of inertia changes at the same rate regardless of the configuration considered. Core configurations tend to significantly decrease, together with the maximum mass, the maximum moment of inertia allowed for stable configurations. Halos similar to before follows closely the pure BM case, only slightly decreasing the values of $I_\BM$ allowed for stable configurations. Increasing the DM content from $f_\DM = 5\%$ to $f_\DM = 15\%$ results in a deviation from the BM case for a given mass, effectively decreasing the maximum $I_\BM$ allowed.

\subsection{Rotating dark matter}\label{sec:resultsCounter}
In the previous subsection, we considered DMANSs configurations with a rotating BM component and a non-rotating DM component. This approach allowed us to establish a baseline for understanding how DM impacts the structure of NSs without contributing directly to the rotational dynamics. Here, we lift the assumption of non-rotating DM, enabling both components to rotate independently.

\begin{figure}
\centering
\includegraphics[width=0.95\columnwidth]{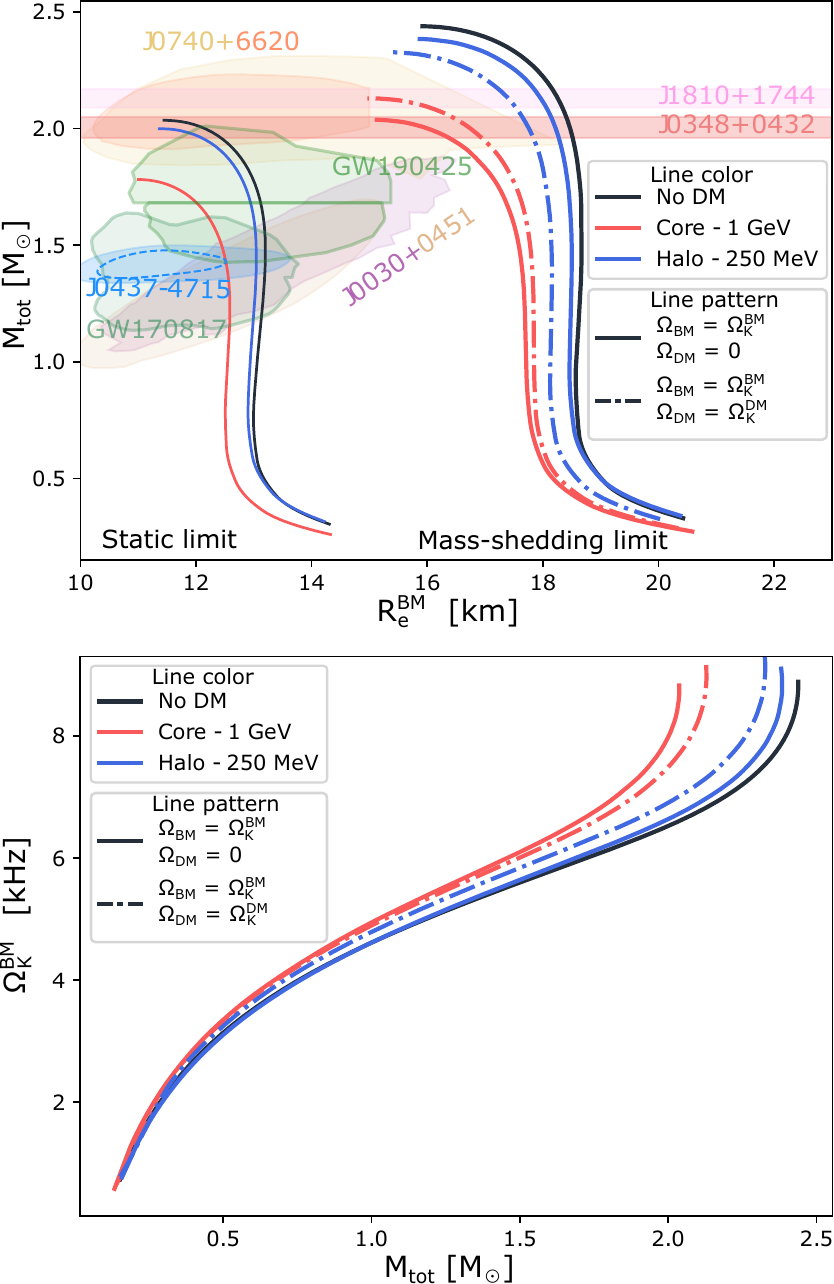}
\vspace*{-1.5ex}
\caption{\textit{Top}: Comparison between the M-R curves of a maximally rotating NS with non-rotating DM (solid lines) and DM co-rotating at its Keplerian angular velocity (dash-dotted lines), $\Omega_\DM = \Omega^\DM_\K$. \textit{Bottom:} Same comparison for the NS's maximum angular velocity as a function of the total gravitational mass.} 
\label{fig:fig5}
\end{figure}

\begin{figure*}
    \centering
    \includegraphics[width=\textwidth]{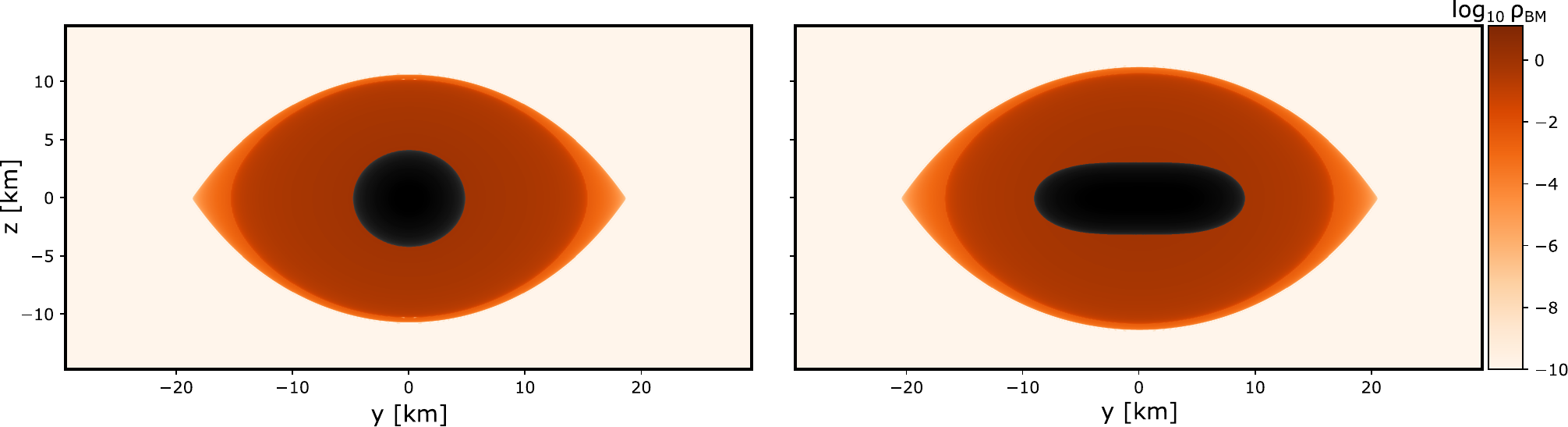}
    \vspace*{-1.5ex}
    \caption{Density profiles in the yz plane at $x = 0$ for a NS rotating at $\Omega^\BM_\K$, $M_\tot = 2\Msun$, $m_\DM = 1\gev$ and $f_\DM = 5\%$. \textit{Left}: core configuration, $\Omega^\BM_\K = 8076 \hz$ and two fluids co-rotating with the same angular velocity $\Omega_\DM = \Omega^\BM_\K$. \textit{Right}: core configuration $\Omega^\BM_\K = 7279 \hz$ and two fluids counter-rotating with the same angular velocity $\Omega_\DM = -\Omega^\BM_\K$.}
    \label{fig:fig6}
\end{figure*}

Fig.~\ref{fig:fig5} reproduces Fig.~\ref{fig:fig2} with the same color scheme and line styles as in the previous subsection; dash-dotted lines represent the configurations characterized by $f_\DM = 5\%$, $\Omega_\BM = \Omega^\BM_\K$ and $\Omega_\DM\ = \Omega^\DM_K$. This is practically the fastest rotation of the DM component that can be imposed and is thus useful to see the maximum difference with the $\Omega_\DM\ = 0$ regime. 

Firstly, it is important to highlight that the blue dash-dotted line in Fig.~\ref{fig:fig5} does not represent a halo of DM surrounding an NS anymore. It is instead a very large core with $R^\DM_e \approx R^\BM_e$. The low-mass halo of DM expands, due to rotation, less than the BM does, ultimately becoming engulfed by the NS. For convenience, we will refer to it as ``halo" nonetheless, to underline the value of the DM particle.

The core configurations increase the maximum mass of the sequence by about 4.5\% while the halo configurations reduce it by about 2.5\%, going from the thick solid lines to the dash-dotted ones. Comparing--where possible-- the equal mass configurations, we notice that rotating cores tend to increase the equatorial radius of the NS while rotating halos tend to significantly decrease it. The Kepler frequency reflects this change: at a constant mass, it decreases for cores and increases for halos.

\begin{figure}
\centering
\includegraphics[width=0.95\columnwidth]{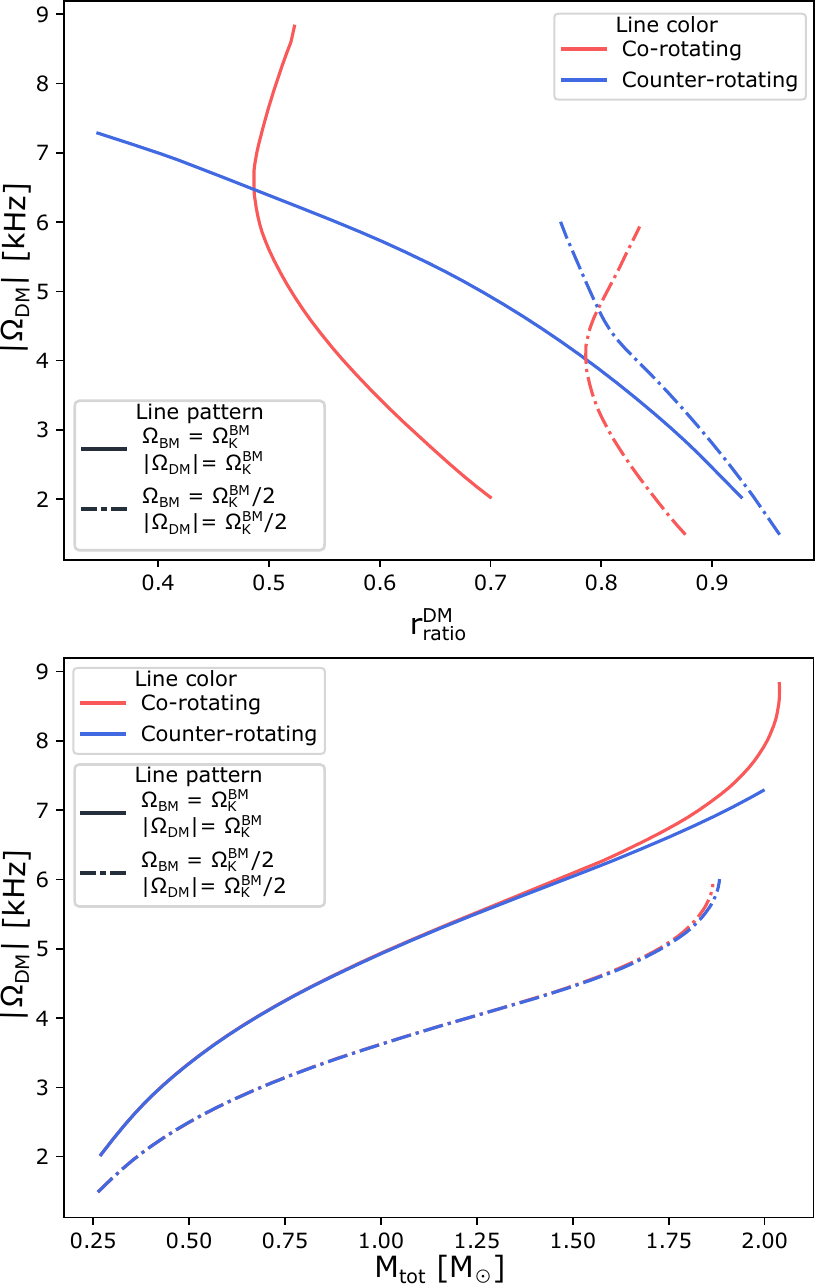}
\vspace*{-1.5ex}
\caption{Comparison between co-rotating (red) and counter-rotating (blue) core configurations for $m_\DM=1\gev$,  $f_\DM = 5\%$ and $\Omega_\BM = \left|\Omega_\DM\right| = \Omega^\BM_\K$ (solid line) or $\Omega_\BM = \left|\Omega_\DM\right| = \Omega^\BM_\K / 2$ (dash-dotted line). \textit{Top}: Relation between the angular velocity $\left|\Omega_\DM\right|$ and the ratio of polar to equatorial radius $\rrDM$ of the corresponding configuration. \textit{Bottom:} Relation between the angular velocity $\left|\Omega_\DM\right|$ and the total mass of the system.} 
\label{fig:fig7}
\end{figure}

Lastly, Fig.~\ref{fig:fig6} compares two core configurations with $M_\tot = 2\Msun$, $f_\DM = 5\%$, and $\left|\Omega_\DM\right| = \Omega_\BM = \Omega^\BM_\K$. 
In the left panel, both fluids co-rotate with the same angular velocity ($\Omega_\DM = \Omega^\BM_\K > 0$), representing a system that has synchronized over time and reached dynamic equilibrium. In this case, the deformation of the DM component remains moderate, with the mass-radius relation showing only minor deviations from the corresponding non-rotating case.

In the right panel of Fig.~\ref{fig:fig6}, the two fluids are instead counter-rotating with $\Omega_\DM = - \left|\Omega^\BM_\K\right|$. A notable difference in this configuration is the deformation of the DM component. While in the co-rotating case, the ratio of polar to equatorial radius is approximately $\rrDM \approx 0.51$, in the counter-rotating case it decreases significantly to $\rrDM \approx 0.34$. 

The interplay between the metric fields in Eq.~\eqref{eq:metric}, the fluid angular velocity Eq.~\eqref{eq:Omegax} and the Kepler angular velocity Eq.~\eqref{eq:Ok} is complex. Significant insight is given by Fig.~\ref{fig:fig7}: in the top panel we report the relation between the absolute value of the DM angular velocity for co-rotating (red) and counter-rotating (blue) core configurations; the bottom panel shows instead the dependence of $\Omega_\DM$ on the total gravitational mass. We use solid lines for $\Omega_\BM = \left|\Omega_\DM\right| = \Omega^\BM_\K$ and dash-dotted lines for $\Omega_\BM = \left|\Omega_\DM\right| = \Omega^\BM_\K / 2$.

For both values of the angular velocity, low mass systems behave as expected: increasing the required angular velocity (i.e., decreasing the input parameter $\rrBM$) the deformation of the DM fluid has to increase (i.e., $\rrDM$ has to decrease). It is unexpected to notice however that co-rotating configurations can only achieve the desired angular velocity for much smaller values of $\rrDM$ with respect to counter-rotating ones. As the total mass and angular velocity increase, counter-rotating configurations favor more deformed systems; on the other hand, the co-rotating sequence at some point inverts this trend and starts increasing the angular velocity while also decreasing the deformation (moving towards higher values of $\rrDM$). The DM is ``pushed" by the frame-dragging caused by the BM when they co-rotate. To reach high angular velocities a lower relative angular velocity $\Omega_\DM - \omega$ is therefore necessary, which is reflected in a smaller deformation from spherical symmetry. The opposite situation occurs for counter-rotating configurations, that move against frame-dragging.

In Fig.~\ref{fig:fig7} the line for the counter-rotating configuration ends before it reaches its last stable configuration. This is most probably a numerical issue related to the difficulty of obtaining solutions having extreme deformation with the \texttt{RNS} code.

\subsection{Effect of the variation of the BM EoS}
\label{sec:resultsDD2noY}

\begin{figure}
\centering
\includegraphics[width=0.95\columnwidth]{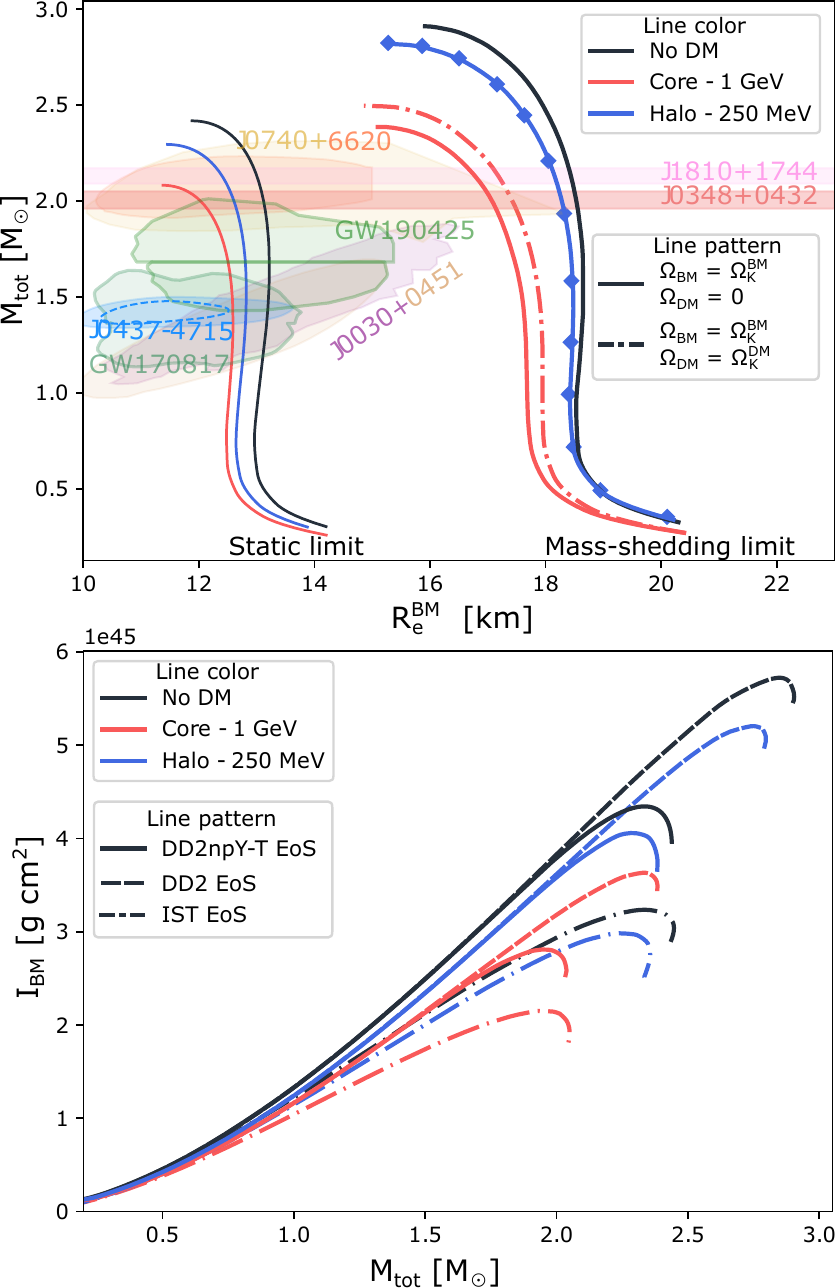}
\vspace*{-1.5ex}
\caption{\textit{Top:} M-R diagram for the DD2 EoS. Thin lines represent static configurations for reference, while thick solid lines represent maximally rotating NSs with non-rotating DM. Dash-dotted lines represent sequences with $\Omega_\BM = \Omega^\BM_\K$ and $\Omega_\DM=\Omega^\DM_\K$. With red (blue) we indicate the curves that characterize the core (halo) configuration with particle mass of 1 GeV (250 MeV) at constant DM fraction $f_\DM = 5\%$. Since halo configurations with a non-rotating DM and DM rotating at its Kepler limit cannot be visually distinguished, we plot the latter with diamonds appearing on top. \textit{Bottom:} Comparison of the moment of inertia computed for three EoSs for $\Omega_\BM = \Omega^\BM_\K$ and $\Omega_\DM=0$ at $f_\DM = 5\%$.} 
\label{fig:fig8}
\end{figure}

As the performed analysis incorporates two underlying EoSs, for BM and DM, their choice would introduce some uncertainties, while qualitatively the conclusions will remain the same. To address the uncertainties of the dense BM EoS we now turn our attention to the DD2 EoS introduced in Section~\ref{sec:BMEOS}. The top panel of Fig.~\ref{fig:fig8} presents the M-R diagram, where the thin solid lines denote static configurations, thick solid lines correspond to maximally rotating BM with non-rotating DM, and dash-dotted lines represent both fluids at their respective maximum angular velocities. Colors once again distinguish between core and halo configurations.

In the absence of DM, the DD2 EoS, in comparison to the DD2npY-T EoS, allows a significantly higher maximum mass, around $2.41\Msun$. As a result, even static core configurations with $f_\DM = 5\%$ can satisfy the observational data on heavy pulsars: PSR J0348+0432 $M = 2.01\pm0.04\Msun$~\cite{Antoniadis:2013pzd} and barely fall below the $M = 2.13\pm0.04\Msun$ of PSR J1810+1744~\cite{Romani:2021xmb}. Introducing DM rotation in core configurations has the same effect that could be seen in Fig.~\ref{fig:fig5}: the maximum allowed mass increases and the corresponding radius slightly decreases. On the other hand, halo configurations are almost identical whether or not the DM fluid rotates. This is because the extended halo surrounding the NS, 
with an equatorial radius much larger than that of the BM, has a significantly lower Kepler frequency (approximately $2\khz$) than that of the baryonic component (approximately $7\khz$). Consequently, the effect of DM rotation is negligible.

Lastly, we show how the moment of inertia depends on the choice of the BM EoS. To provide a more comprehensive analysis, in addition to the stiffer DD2npY-T and DD2 EoSs we also consider a softer IST EoS. The sequences are reported in the bottom panel of Fig.~\ref{fig:fig8} with, respectively, solid, dashed, and dash-dotted lines. The values are computed at $f_\DM = 5\%$ for maximally rotating NSs and non-rotating DM.

In general, softer BM EoSs correspond to a smaller moment of inertia~\cite{Lattimer:2004nj}. A similar effect can be achieved with the accumulation of heavy DM in the NS core, whereas light DM forming an extended halo increases the moment of inertia. Additionally, we present a transition case where a halo with a low DM fraction also leads to a decrease in the moment of inertia.

\section{Summary and conclusions}
\label{sec:conclusions}

Due to their fast spin and long evolution history, millisecond pulsars are ideal laboratories to study the effect of DM on NSs. In addition, millisecond pulsars representing old NSs are expected to contain a higher DM fraction, particularly important for this analysis. To allow the study of rapidly rotating DM-admixed NSs we have modified the public code \texttt{RNS} to include a second fluid. We have assumed the two fluids interact only gravitationally, are cold, and rotate rigidly around a shared axis. The DM distribution is automatically determined by the corresponding EoS and the central energy density value. The rotation is considered to be freely defined in both magnitude and direction. 

DM is modeled by a bosonic self-interacting fluid with a quartic repulsive potential. To describe BM we have utilized three EoSs representing different classes of models with varying particle compositions, stiffness, and nuclear matter properties at the saturation density. Particularly, the DD2npY-T and DD2 EoSs with and without hyperons, respectively, formulated within the relativistic density functional approach and the parametric IST EoS. \texttt{RNS} is however not limited to these choices, and other EoSs can be studied as simply as they can be tabulated.

Since this is a multi-parameter system, we have chosen to present sequences with a fixed DM fraction while considering two cases for the BM angular frequency: one corresponding to the Keplerian limit, where the star rotates at its maximum allowed frequency, and another for a realistic millisecond pulsar rotating at $716\hz$, the fastest observed spin rate to date~\cite{Hessels:2006ze,Jaisawal_2024}.

Our results indicate that incorporating DM in NSs preserves the overall qualitative behavior observed for pure baryonic stars. However, the quantitative changes depend strongly on the chosen parameters for both BM and DM components, with the DM particle mass and self-interaction playing a key role in distinguishing the core from the halo configurations in the static limit. Introducing BM rotation may blur this distinction—e.g., a halo may shift to a mixed state where the BM equatorial radius exceeds the DM's, but the DM's polar radius remains larger. Notably, at high masses, we observe a degeneracy of the oblateness among different DM configurations (at fixed baryonic radius and DM fraction). The oblateness peaks and then decreases linearly with the baryonic equatorial radius with nearly identical slopes for different DM particle masses. The moment of inertia also shows significant changes, decreasing for both core and halo configurations relative to the pure baryonic case.

The impact of rotation of the DM component at its Keplerian limit depends on its configuration: cores increase the maximum mass--as rotation provides an apparent force that counteracts gravity-- while halo configurations are highly sensitive to the parameters of both EoSs. In some cases, the DM remains as an extended halo with a minimal effect on the mass-radius diagram. This is due to the low Kepler frequency at which the DM component can rotate when compared to BM. In other cases DM becomes engulfed by the NS, effectively forming a very large core.

Finally, we examined consistently the case when DM rotates with the same absolute value of the angular velocity as BM, either co-rotating or counter-rotating, reaching the BM Keplerian limit. An interesting observation for the core DM configurations is the following. For low-mass systems, where the BM Kepler angular velocity is low, configurations with co-rotating DM need a larger deformation of the DM component than counter-rotating DM. As the mass increases and frame-dragging effects grow, co-rotating configurations achieve higher angular velocities with a smaller deformation. In contrast, counter-rotating configurations require progressively larger deformations to reach the BM Kepler limit, reaching very oblate DM configurations with a ratio of the polar to the equatorial DM radius lower than 0.35.

This analysis can be further extended in various ways. In particular, it would be interesting to investigate how the presented results depend on the variation in the self-interaction strength $\lambda$ of the DM field, as described by Eq.~\eqref{eq:DMLagr}. Comparison with experimental data could exclude portions of the $(\lambda, m_\DM)$ parameter space. This can already be glimpsed for the DD2npY-T EoS looking at our results where, for the chosen $\lambda$ and DM fraction, static core configurations with $m_\DM = 1\gev$ are below the $2\Msun$ measurement~\cite{Miller:2021qha, Riley:2021pdl}. Maximally rotating stars would instead be driven below the same threshold for higher values of the mass. Another possibility is to move beyond the assumption of uniform rotation—suitable for old, isolated stars—and explore differential rotation, which is more relevant for remnants of binary neutron star mergers~\cite{Camelio:2019rsz,Camelio:2020mdi,Iosif:2021aum,Iosif:2021qlv,Cassing:2024dxp,Cipriani:2024bcc}.

The physical mechanisms driving dark matter rotation in admixed systems remain an open question, with several intriguing scenarios proposed. These include the inheritance of angular momentum from preexisting structures, asymmetric accretion as the neutron star moves through the surrounding dark matter halo (similar to the proposed seasonal modulation of dark matter flux due to the Earth's motion relative to the galactic halo~\cite{PhysRevD.37.1353}) and the aftermath of dark matter-admixed binary neutron star mergers~\cite{Giangrandi:2025rko}.

The analysis performed in this work has significant implications for both current and future radio and X-ray observations of compact stars. The Square Kilometre Array (SKA) radio telescope~\cite{Watts:2014tja}, along with other radio telescopes, will focus on precise measurements of NS moment of inertia, pulsar timing, and other NS properties. However, it is crucial to understand how accumulated DM might affect these quantities. As demonstrated in this study, the M(R) relations and moment of inertia exhibit a high degree of degeneracy, making it challenging to disentangle the effects of DM from the BM EoS. Therefore, measuring only the NS radius at a given mass would not be sufficient to distinguish DM effects or accurately probe the dense matter EoS.

On the other hand, present and future X-ray missions, such as NICER, eXTP (Enhanced X-ray Timing and Polarimetry Mission), and NewAthena (New Advanced Telescope for High-Energy Astrophysics), may encounter another effect related to DM -- a potential impact on the formation and observation of hotspots~\cite{Rutherford:2022xeb, Shakeri:2022dwg}.

The presence of dark matter in neutron stars could be probed with static and dynamic effects. They include measuring the mass and radius of neutron stars with a few-percent accuracy~\cite{Giangrandi:2022wht}, modification of the pulsar pulse profile due to the extra light-bending~\cite{2022ApJ...936...69M} and/or gravitational microlensing in the case of the existence of a dark halo, modification of the cooling rate of compact stars~\cite{10.1093/mnras/stae337} as well as the aforementioned dark matter-admixed binary neutron star systems~\cite{Giangrandi:2025rko}. This is the subject of ongoing research.

In conclusion, this analysis highlights that interpreting X-ray, radio, and gravitational wave observations without considering the presence of accumulated DM could lead to missing crucial information or even misinterpreting the properties of strongly interacting matter at high densities. This becomes particularly relevant for NSs located in the Galactic center or dense DM regions, where DM density is expected to be high.

\acknowledgments
The authors thank Fabrizio Nesti, Massimo Mannarelli and Nils Andersson for their useful comments and discussions. The work of E.G. was supported by national funds from FCT - Fundação para a Ciência e Tecnologia, I.P. through the projects UIDB/04564/2020 and UIDP/04564/2020, with DOI identifiers 10.54499/UIDB/04564/2020 and 10.54499/UIDP/04564/2020, respectively. E.G. also acknowledges the support from Project No. PRT/BD/152267/2021. V.S. gratefully acknowledges support from the UKRI-funded ``The next-generation gravitational-wave observatory network'' project (Grant No. ST/Y004248/1). S.Y. is supported by the European Union-NextGenerationEU, through the National Recovery and Resilience Plan of the Republic of Bulgaria, project No. BG-RRP-2.004-0008-C01. D.D.  acknowledges financial support via an Emmy Noether Research Group funded by the German Research Foundation (DFG) under grant no. DO 1771/1-1.

\appendix
\section{Implementation details of the \texttt{RNS} code}\label{app:implem}
\subsection{Model inputs and computational parameters}
The algorithm outlined in Section~\ref{sec:NumFrameStructEq} requires the specification of input parameters for \texttt{RNS}. Both metric and matter fields are fully characterized by two values: the central energy density, $\varepsilon^\X_c$, and the desired ratio between the polar and equatorial radii, $r^{\X}_\text{ratio}$. Note that this ratio is imposed not for the circumferential radius \eqref{eq:circRad}, but for the radii in the coordinates of metric~\eqref{eq:metric}.

Other physical quantities, e.g., mass, DM fraction, or angular velocity, can also be specified. However, these models are computed iteratively by adjusting the appropriate fundamental parameter until a configuration is found within a specified tolerance. The relative error in determining these parameters is set to $10^{-3}$ by default. This should not be confused with the numerical tolerance required for the convergence of the equilibrium equations' solution, which is set to $10^{-10}$. The general algorithm used goes as follows.

\begin{itemize}
    \item Initial conditions are computed solving the TOV equations \eqref{eq:TOV} for some central energy densities $\varepsilon^\X_c$, taken as first guesses.
    \item The four free parameters $\varepsilon^\BM_c$, $\varepsilon^\DM_c$, $\rrBM$, and $\rrDM$ can either be given as input or fixed with four nested root finding loops to achieve some desired property of the computed system. For instance, $\varepsilon^\BM_c$ is changed to find constant mass sequences, $\varepsilon^\DM_c$ to fix the DM fraction, $\rrBM$ and $\rrDM$ allow to achieve the target angular velocity for the respective fluid. For each iteration of the innermost root finding, the equilibrium equations are solved, and the properties of the system are computed.
    \item Once convergence is achieved, the final model with desired parameters is recomputed, and key stellar properties, including mass, radius, and angular momentum, are calculated.
\end{itemize}
It is worth mentioning that algorithms for finding roots in multidimensional spaces do exist, but their implementation here is not straightforward. Instead, the algorithm described navigates the parameter space in small, incremental steps. This approach ensures that the initial guess is sufficiently close to the solution, guaranteeing convergence. More sophisticated methods, such as Broyden's algorithm~\cite{Broyden1965}, are available but typically involve larger steps through the parameter space, at least in the beginning. These larger steps often result in failures when solving the equilibrium equations, making them less reliable in this context.

\subsection{Einstein field equations for rotating systems}
Einstein's field equations
\begin{equation}
    G_{\mu\nu} = 8\pi T_{\mu\nu}
\end{equation}
need to be solved to specify the metric~\eqref{eq:metric}. Following~\cite{1971ApJ...167..359B} and replacing $T_{\mu\nu}$ with Eq.~\eqref{eq:tmunu}, it is possible to compute the non-trivial fields equations from the tt, t$\phi$, and $\phi\phi$ components. Defining $\mu=\cos\theta$ to avoid the singularity in the pole of $\cot{\theta}$, the equations for the metric fields $\rho$, $\gamma$ and $\omega$ are
\begin{subequations}
\begin{align}
\Delta \left(\rho \,\ee^{\gamma / 2}\right) = S_\rho\, ,\\
\left( \Delta + \frac{1}{r} \frac{\partial}{\partial_r} - \frac{1}{r^2}\mu\frac{\partial}{\partial\mu} \right)\gamma \,\ee^{\gamma /2} = S_\gamma\, ,\\
\left( \Delta + \frac{2}{r} \frac{\partial}{\partial_r} - \frac{2}{r^2}\mu\frac{\partial}{\partial\mu} \right)\omega \,\ee^{(\gamma - 2 \rho)/2} = S_\omega\, 
\end{align}
\end{subequations}
with $\Delta$ the Laplacian operator in spherical coordinates and 
\begin{widetext}
\begin{subequations} \label{eq:A3}
\begin{align}
\begin{split}
S_\rho(r, \mu) = ~&e^{\frac{\gamma}{2}} \left\{ 8\pi \ee^{2\alpha} \left( \varepsilon_\BM + P_\BM \right) \frac{1 + v_\BM^2}{1 - v_\BM^2} + 8\pi \ee^{2\alpha} \left( \varepsilon_\DM + P_\DM \right) \frac{1 + v_\DM^2}{1 - v_\DM^2} +\right.\\
&r^2(1 - \mu^2)\ee^{-2\rho} \left[ \left(\partial_r \omega\right)^2 + \frac{1}{r^2}(1 - \mu^2)\left(\partial_\mu \omega\right)^2 \right] + \frac{\partial_r \gamma}{r} - \frac{1}{r^2} \mu \partial_\mu \gamma +\\
&\left.\frac{\rho}{2} \left[ 16 \pi e^{2\alpha} \left( P_\DM + P_\BM \right)
- \partial_r \gamma \left( \frac{1}{2} \partial_r\gamma + \frac{1}{r} \right) - \frac{\partial_\mu \gamma}{r^2} \left( \frac{1}{2} \partial_\mu\gamma(1 - \mu^2) - \mu \right) \right] \right\}\, , 
\end{split} \label{eq:A3a} \\
\begin{split}
 S_\gamma(r, \mu) = ~&\ee^{\frac{\gamma}{2}} \left\{ 16\pi\ee^{2\alpha} (P_\DM + P_\BM) + \frac{\gamma}{2}\left[ 16\pi\ee^{2\alpha} (P_\DM + P_\BM) - \frac{\left(\partial_r \gamma\right)^2}{2} - \frac{1}{2r^2}(1 - \mu^2) \left(\partial_\mu\gamma\right)^2 \right]\right\}\, ,
\end{split} \label{eq:A3b} \\
\begin{split}
S_\omega(r, \mu) =~&\ee^{\frac{\gamma - 2\rho}{2}} \left\{-16\pi\ee^{2\alpha} \frac{(\Omega_\BM - \omega)(\varepsilon_\BM + P_\BM)}{1 - v_\BM^2} -16\pi\ee^{2\alpha} \frac{(\Omega_\DM - \omega)(\varepsilon_\DM + P_\DM)}{1 - v_\DM^2} +\right.\\
&\omega \left[ -8\pi\ee^{2\alpha} \frac{(1 + v_\BM^2)\varepsilon_\BM + 2 v_\BM^2 P_\BM}{1 -  v_\BM^2} -8\pi\ee^{2\alpha} \frac{(1 + v_\DM^2)\varepsilon_\DM + 2 v_\DM^2 P_\DM}{1 -  v_\DM^2} - \right.\\
&\frac{1}{r}\left(2\partial_r \rho + \frac{1}{2}\partial_r\gamma \right) + \frac{1}{r^2} \mu \left( 2\partial_\mu \rho + \frac{1}{2}\partial_\mu\gamma \right) + \frac{1}{4}\left(4(\partial_r\rho)^2 - (\partial_r\gamma)^2\right) +\\
&\left.\left.\frac{1}{4 r^2}(1-\mu^2)\left(4(\partial_\mu\rho)^2 - (\partial_\mu\gamma)^2\right) - r^2 (1-\mu^2) \ee^{-2\rho} \left( \left(\partial_r\omega\right)^2 + \frac{1}{r^2}(1-\mu^2)\left(\partial_\mu \omega\right)^2 \right)\right]\right\}\, .
\end{split} \label{eq:A3c}
\end{align}
\end{subequations}
\end{widetext}
The method for solving these equations, detailed in~\cite{Komatsu:1989zz}, involves converting them into integral equations using multi-dimensional Green's functions.
In particular, we obtain
\begin{equation}
\rho = -\frac{\ee^{-\gamma/2}}{4\pi} \int_0^\infty\!\!\!\!\!\dd \tilde{r}\int_{-1}^1\!\!\!\!\!\dd \tilde{\mu}\int_0^{2\pi}\!\!\!\!\!\dd \tilde{\phi}\, \frac{\tilde{r}^2}{\left| \mathbf{r} - \mathbf{\tilde{r}} \right|} S_\rho\, ,
\end{equation}
\begin{equation}
r\sin\theta\,\gamma = \frac{\ee^{-\gamma / 2}}{2\pi} \int_0^\infty\!\!\!\!\!\dd \tilde{r}\int_{0}^{2\pi}\!\!\!\!\!\dd \tilde{\theta}\,\tilde{r}^2 \sin\tilde{\theta}\log \left| \mathbf{r} - \mathbf{\tilde{r}} \right| S_\gamma\, ,
\end{equation}
and finally
\begin{equation}
\begin{split}
r\sin\theta\cos\phi\,\omega &= -\frac{\ee^{(2\rho -\gamma) / 2}}{4\pi} \times\\
&\int_0^\infty\!\!\!\!\!\dd \tilde{r}\int_{0}^{\pi}\!\!\!\!\!\dd \tilde{\theta}\int_0^{2\pi}\!\!\!\!\!\dd \tilde{\phi}\,\tilde{r}^3 \frac{\sin^2\tilde{\theta} \cos \tilde{\phi}}{\left| \mathbf{r} - \mathbf{\tilde{r}} \right|} S_\omega\, .
\end{split}
\end{equation}
This automatically results in the correct asymptotic behavior of the metric fields, i.e. $\rho \sim O\left(\frac{1}{r}\right)$, $\gamma \sim O\left(\frac{1}{r^2}\right)$ and $\omega \sim O\left(\frac{1}{r^3}\right)$. Finally, the potential $\alpha$ is determined from the other metric fields solving Eq.~(36) of~\cite{Komatsu:1989zz}.

\end{document}